\documentclass[useAMS,usenatbib]{mn2e}
\usepackage{epsfig}
\usepackage{amsmath}

\newcommand{\be}{\begin{equation}}
\newcommand{\beq}{\begin{equation}}
\newcommand{\ba}{\begin{eqnarray}}
\newcommand{\ee}{\end{equation}}
\newcommand{\eeq}{\end{equation}}
\newcommand{\ea}{\end{eqnarray}}

\newcommand{\apj}{ApJ}
\newcommand{\apjl}{ApJL}
\newcommand{\mnras}{MNRAS}
\newcommand{\aj}{AJ}
\newcommand{\apjs}{ApJS}

\def\lsim{~\rlap{$<$}{\lower 1.0ex\hbox{$\sim$}}}

\def\gsim{~\rlap{$>$}{\lower 1.0ex\hbox{$\sim$}}}

\title[21cm emission from DLAs]{A Method to Measure the Mass of Damped Ly$\alpha$ Absorber Host Galaxies Using Fluctuations in 21cm Emission}

\author[Wyithe]{J. Stuart B. Wyithe\\
School of Physics, University of Melbourne, Parkville, Victoria,
Australia\\Email: swyithe@unimelb.edu.au}

\begin{document}


\maketitle

\label{firstpage}
\begin{abstract}

Observations of damped Ly$\alpha$ absorbers (DLA) indicate that the
fraction of hydrogen in its neutral form (HI) is significant by mass
at all redshifts. This gas represents the reservoir of material that
is available for star formation at late times. As a result,
observational identification of the systems in which this neutral
hydrogen resides is an important missing ingredient in models of
galaxy formation. Precise identification of DLA host mass via
traditional clustering studies is not practical owing to the small
numbers of known systems being spread across sparsely distributed
sight lines. However following the completion of reionization, 21cm
surface brightness fluctuations will be dominated by neutral hydrogen
in DLAs. No individual DLAs could be detected in 21cm
emission. Rather, observations of these fluctuations will measure the
combined clustering signal from all DLAs within a large volume. We
show that measurement of the spherically averaged power-spectrum of
21cm intensity fluctuations due to DLAs could be used to measure the
galaxy bias for DLA host galaxies when combined with an independent
measurement of the cosmological HI mass density from quasar absorption
studies. Utilising this technique, the low frequency arrays now under
construction could measure the characteristic DLA host mass with a
statistical precision as low as 0.3 dex at $z\ga4$. In addition, high
signal-to-noise observations of the peculiar-motion induced anisotropy
of the power-spectrum would facilitate measurement of both the DLA
host mass and the cosmic HI density directly from 21cm fluctuations.
By exploiting this anisotropy, a second generation of low frequency
arrays with an order of magnitude increase in collecting area could
measure the values of cosmic HI density and DLA host mass, with
uncertainties of a few percent and a few tens of percent respectively.

\end{abstract}


\section{Introduction}

The primary avenue for study of the physical properties of the
inter-galactic medium utilises the Ly$\alpha$ absorption systems
observed along the lines-of-sight towards high redshift quasars. These
absorbing systems include the Ly$\alpha$ forest (with column depth
$n<10^{17}$cm$^{-2}$), Ly-limit systems
($10^{17}$cm$^{-2}<n<2\times10^{20}$cm$^{-2}$), and damped Ly$\alpha$
absorbers (DLA, $n>2\times10^{20}$cm$^{-2}$). The latter are
self-shielded and have been shown to host $>80\%$ of the neutral
hydrogen (HI) gas during most of cosmic time $z\la5$ (Prochaska et
al.~2005).

As the primary reservoirs of neutral gas, it is thought that DLAs
provide the dominant sites for star formation, and represent the
progenitors of modern galaxies (Woolfe, Gawiser \& Prochaska~2005).
Thus the identity of DLA hosts is crucial for our understanding of
galaxy formation, and of the buildup of stellar mass in the
Universe. DLA hosts have been identified in a few cases (e.g. Colbert
\& Malkan~2002; Warren et al.~2001), and observations of these (with
some exceptions) are consistent with their being at the faint end of
the Lyman break galaxy (LBG) population. On the other hand, these
studies will more easily discover hosts in the most luminous systems,
and could therefore be systematically biased in their
conclusions. Within the cold-dark-matter model the masses of objects
may be determined independently from their luminosity by studying
their clustering properties and appealing to theoretical predictions
of the bias relative to the underlying mass distribution. At high
redshift $z>3$, clustering analyses have been used in this way to
study the host halo masses of quasars (Shen et al.~2007; White,
Martini \& Cohn~2007), and LBGs (Adelberger et al.~2005). However the
numbers of DLAs known is small, occurring only once in every few
quasar spectra, and so no sample of sufficient size to allow for an
auto-correlation clustering analysis exists (Cooke et al.~2006). As a
result a quantitative measurement of DLA host mass has remained
illusive. Some progress has been made through comparison of clustering
with the LBG population through cross-correlation. Cooke et al.~(2006)
have performed a survey for LBGs in regions surrounding known DLA
systems. This sample allows for measurement of the cross-correlation
between DLAs and LBGs. Cook et al.~(2006) find the cross-correlation
length to be comparable to the correlation length (Adelberger et
al.~2005) among the large LBG sample of Steidel et al.~(2003), and so
conclude that DLA systems have similar masses to LBGs. They argue for
a host halo mass of $10^{9}M_\odot\la M\la10^{12}M_\odot$. This
interpretation is consistent with modeling studies carried out in a
cosmological framework (Nagamine et al.~2007).

Alternatively, the presence of cold gas is also traced by MgII
absorption. Indeed, MgII absorbers have been shown to be associated
with neutral hydrogen absorbers over a range of column densities,
including DLAs (e.g. Rao et al.~2006), and so MgII absorption is used
as a proxy to study the HI galaxy population at $z\la1.5$.  The host
halo masses of MgII absorbers have been estimated at $z\sim0.5$ via
cross-correlation with luminous red galaxies in the Sloan Digital Sky
Survey Data Release 3 (Bouch{\'e} et al.~2006), yielding a host mass
of $M\sim10^{12}M_\odot$. The large number of objects available (both
MgII absorbers and galaxies) yields a statistical accuracy of a factor
of 2 in host halo mass, much better than is available using comparable
techniques for DLAs at higher redshift (Cook et al.~2006). In
addition, host masses have been estimated via direct kinematic
measurement in spatially resolved followup spectroscopy (Bouch{\'e} et
al.~2007), yielding estimates that are consistent with clustering
results.

The small number of known DLAs is due to the low density of bright
background sources. An alternative approach is therefore to observe
the DLA population directly through 21cm emission. There is an
extensive literature describing the use of redshifted 21cm
observations as a probe of the process of reionization in the high
redshift IGM [see Furlanetto, Oh \& Briggs~(2006) for an extensive
discussion].  Reionization starts with ionized (HII) regions around
galaxies, which later grow to surround groups of galaxies. The process
of reionization is completed when these HII regions overlap (defining
the so-called {\it overlap} epoch) and fill-up most of the volume
between galaxies.  The conventional wisdom has been that the 21cm
signal disappears after the {\it overlap} epoch, because there is
little neutral hydrogen left through most of intergalactic
space. However, the following simple estimate can be used to
demonstrate that 21cm emission should be significant even after
reionization is completed (Wyithe \& Loeb~2008). As mentioned above,
DLA systems are believed to contain the majority of HI at high
redshifts. Indeed, observations of DLAs out to a redshift of $z\sim4$
show the cosmological density parameter of HI to be $\Omega_{\rm
HI}\sim10^{-3}$ (Prochaska et al ~2005). In the standard cosmological
model the density parameter of baryons is $\Omega_{\rm b}\sim 0.04$,
so that the mass-averaged neutral hydrogen fraction at $z\sim4$ (long
after the end of the HII overlap epoch) is $x_{\rm
HI}\equiv\Omega_{\rm HI}/\Omega_{\rm b}\sim0.02$.

At $z\sim4$, the brightness temperature contrast of redshifted 21cm
emission will be $\Delta T\sim0.5$mK. Moreover on the $R\sim10$
co-moving Mpc scales relevant for upcoming 21cm experiments, the {\it
root-mean-square} amplitude of density fluctuations at $z\sim4$ is
$\sigma\sim0.2$. Hence, we expect fluctuations in the 21cm intensity
field due to DLAs to be at least $\sim0.1$mK on 10~comoving Mpc
scales, with a boost in this signal if the DLAs are hosted by biased
(i.e. massive) dark matter halos (Wyithe \& Loeb~2008).  The
fluctuations in the 21cm emission signal after reionization are
therefore expected to be only an order of magnitude or so smaller than
the largest fluctuations predicted at any time during the entire
reionization era (e.g. Wyithe \& Morales~2007). Moreover the sky
temperature, which provides the limiting factor in the system noise at
the low frequencies relevant for 21cm studies, is proportional to
$(1+z)^{2.6}$, and so is a factor of $\sim 3.4 [(1+z)/5]^{2.6}$
smaller at low-redshifts than for observations at $z\sim7$.

Several experiments are currently under development which aim to
detect the 21cm signal during reionization (including
MWA~\footnote{http://www.haystack.mit.edu/ast/arrays/mwa/},
LOFAR~\footnote{http://www.lofar.org/}, PAPER
~\footnote{http://astro.berkeley.edu/~dbacker/EoR/},
21CMA~\footnote{http://web.phys.cmu.edu/~past/}) and more ambitious
designs are being planned
(SKA~\footnote{http://www.skatelescope.org/}). In addition, the first
statistical detection of 21cm fluctuations due to discrete, unresolved
clumps of neutral gas was recently made (Pen et al.~2008) through
cross-correlation of the HIPASS (Barnes et al.~2001) 21cm observations
of the local universe with galaxies in the 6 degree field galaxy
redshift survey (Jones et al.~2004; 2005). This detection represents
an important step towards using 21cm surface brightness fluctuations
to probe the neutral gas distribution in the IGM (both during and
after reionization), as well as the mass power-spectrum (PS) and
cosmology (McQuinn et al.~2006; Wyithe, Loeb \& Geil~2008; Chang et
al.~2007; Mao et al.~2008; Loeb \& Wyithe~2008; Pritchard \& Loeb~2008;
Wyithe~2008).

The first generation of low frequency radio arrays will have low
spatial resolution, and insufficient sensitivity to detect individual
DLAs in emission. On the other hand, the redshifted 21cm emission is
sensitive to the total (mass-weighted) optical depth of this neutral
gas. Observations of the redshifted 21cm signal would therefore detect
the total neutral hydrogen content in a volume of IGM dictated by the
observatory beam and frequency band-pass, and as a result each beam
will contain the combined emission from a large number of DLAs. It is
important to note that observation of DLAs via 21cm surface brightness
fluctuations would be fundamentally different from quasar absorption
line studies. The surface brightness fluctuations would be sensitive
to the statistical properties of the population (particularly the
clustering), but could not be used to study individual DLAs.

Although the 21cm emission after HII overlap is dominated by dense
clumps of gas rather than by diffuse gas in the IGM as is the case
before reionization is complete, we do not expect 21cm self absorption
to impact the level of 21cm emission. This conclusion is based on 21cm
absorption studies towards DLAs at a range of redshifts between
$z\sim0$ and $z\sim3.4$, which show optical depths to absorption of
the back-ground quasar flux with values less than a few percent
(Kanekar \& Chengalur 2003; Curran et al. 2007).  The small optical
depth for self absorption is also supported by theoretical
calculations of the 21cm optical depth of neutral gas in high redshift
mini-halos (Furlanetto \& Loeb~2002).  Moreover, DLAs have a spin
temperature that is large relative to the temperature of the cosmic
microwave background radiation, and will therefore have a level of
emission that is independent of the kinetic gas temperature
(e.g. Kanekar \& Chengalur 2003). These factors combine to make
prediction of the 21cm signal from DLAs robust against poorly
understood astrophysical details of galaxy formation.

In this paper we consider the PS of 21cm intensity fluctuations due to
DLA systems at $2.5\la z\la5.5$ where there are complementary quasar
absorption line studies (\S~\ref{secDLAPS}). We show that the
clustering could be accurately determined using low-frequency radio
telescopes currently under construction (\S~\ref{sensitivity}), and
that this clustering will allow for an accurate measurement of the DLA
mass (\S~\ref{DLA}-\ref{PSmu}). We summarise our conclusions in
\S~\ref{conclusion}. Throughout the paper we adopt the set of
cosmological parameters based in part of data from {\it WMAP3}
(Spergel et al. 2007) for a flat $\Lambda$CDM universe.

\section{The 21cm power-spectrum of DLAs}
\label{secDLAPS}

\begin{figure*}
\includegraphics[width=15cm]{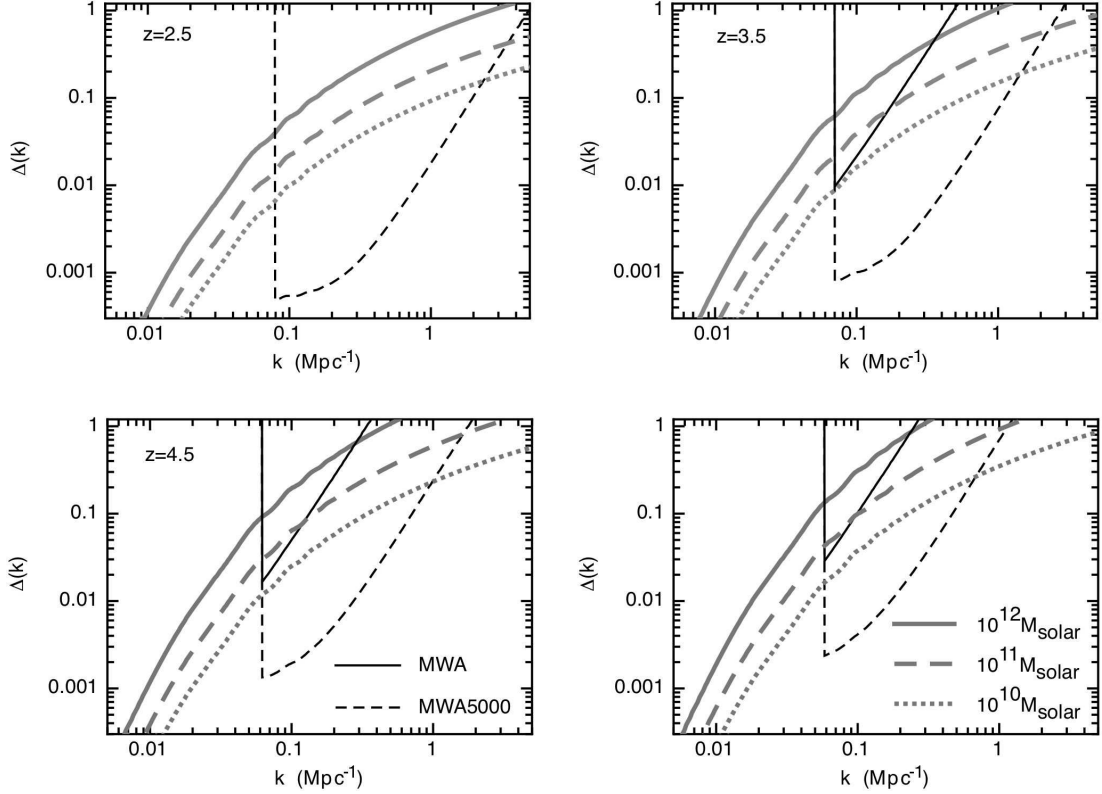} 
\caption{The spherically averaged PS of 21cm fluctuations due to DLAs after reionization (thick
grey lines). We have plotted the PS in dimensionless form $\Delta\equiv k^3P(k)/(2\pi^2)$. The model
PS shown was computed assuming three values for the DLA mass, $M_{\langle b\rangle}=10^{10}M_\odot$, $M_{\langle b\rangle}=10^{11}M_\odot$ and $M_{\langle b\rangle}=10^{12}M_\odot$.  Also shown
for comparison are estimates of the noise for MWA (thin solid lines) and a
future extension MWA5000 (thin dashed lines). In each case we plot the uncertainty within $k$-space bins of size $\Delta k=k/10$, and have assumed observations for 1000hrs in each of 3 different fields. The uncertainty includes a minimum $k$ cutoff due to foreground
subtraction. At $z<3.5$ only the MWA5000 is considered since the MWA antennae are not effective at frequencies greater than $\sim300$MHz. The noise curves, which vary at large $k$ owing to Poisson noise, are shown only for $M_{\langle b\rangle}=10^{12}M_\odot$. }
\label{fig1}
\end{figure*}  

We begin by describing the 21cm PS following the completion of
reionization.  The amplitude of fluctuations in the 21cm signal
depends on the mass-averaged neutral hydrogen fraction in the IGM, and
on the clustering bias of DLA systems. The latter is related to the
galaxy bias of the DLA host dark matter halos. If more massive halos
tend to house more neutral hydrogen, we would expect there to be a
mean relation between DLA column density and the host halo
mass. However DLA systems have a range of column densities, and so we
would also expect significant scatter around this mean relation, both
because of variation in the HI content from galaxy to galaxy, and
because the column density for a particular galaxy will depend on the
line-of-sight probed by a particular observation. In addition, there
is a small contribution to the neutral hydrogen content of the IGM
from Lyman limit systems and the Ly$\alpha$ forest.

For a DLA host with halo
mass $M$, the galaxy bias $b$ may be approximated using the Press-Schechter~(1974)
formalism (Mo \& White~1996), modified to include non-spherical
collapse (Sheth, Mo \& Tormen~2001)
\begin{eqnarray}
\label{bias}
\nonumber
b(M,z) = 1 &+& \frac{1}{\delta_{\rm c}}\left[\nu^{\prime2}+b\nu^{\prime2(1-c)}\right.\\
&&\hspace{10mm}-\left.\frac{\nu^{\prime2c}/\sqrt{a}}{\nu^{\prime2c}+b(1-c)(1-c/2)}\right],
\end{eqnarray}
where $\nu\equiv {\delta_{\rm c}^2}/{\sigma^2(M)}$,
$\nu^\prime\equiv\sqrt{a}\nu$, $a=0.707$, $b=0.5$ and $c=0.6$. Here
$\sigma(M)$ is the variance of the density field smoothed on a mass
scale $M$ at redshift $z$. This expression yields an accurate
approximation to the halo bias determined from N-body simulations
(Sheth, Mo \& Tormen~2001). 

Our approach to computing the DLA PS is to assume that there is a
probability distribution $p(b)db$ for the neutral hydrogen
mass-weighted clustering bias ($b$) of DLAs relative to the underlying
dark matter.  This distribution has a mean $\langle b\rangle$, and a
characteristic mass $M_{\langle b\rangle}$ where $\langle b\rangle =
b(M_{\langle b\rangle})$. The quantity $\langle b\rangle$ is the
typical galaxy bias of halos hosting neutral hydrogen. Since DLAs
dominate the neutral hydrogen mass, and are thought to reside in
massive, biased halos, we refer to this typical bias in the 21cm PS as
the bias of DLA hosts, and $M_{\langle b\rangle}$ as the DLA host mass
throughout this paper.

Our goal is to discuss the statistical precision with which
$M_{\langle b\rangle}$ can be determined via the 21cm PS. As noted
above, while most neutral hydrogen is located in the DLAs, a fraction
of HI in the IGM is observed in Ly-limit systems with column densities
below $2\times10^{20}$cm$^{-2}$. If these systems are found in the
same host halo population as the DLAs, then $M_{\langle b\rangle}$
will faithfully represent the DLA host mass. On the other hand, if
Ly-limit systems are located in smaller less biased halos, then
equation~(\ref{DLAPSlin}) will lead to an underestimate of the DLA
host mass. We do not expect this underestimate to be significant
because the contribution to the observed PS is weighted by the the HI
mass density squared and the bias squared, with DLA systems dominating
both quantities.

Assuming that the relation
between HI and halo mass is independent of large scale over density
(and hence ionizing background), the 21cm PS due to DLAs (Wyithe~2008) 
 is
\begin{equation}
\label{DLAPSlin}
P_{21}(k) = 400\mbox{mK}\left(\frac{1+z}{7.5}\right) x_{\rm HI}^2P(k)\left[\langle b\rangle+f\mu^2\right]^2,
\end{equation}
where $x_{\rm HI}\equiv\Omega_{\rm HI}/\Omega_{\rm b}$ is the mass
weighted neutral fraction of hydrogen in the Universe.  The PS is
evaluated at a wavenumber $\vec{k}$ with modulus $k$ and
$\mu=\cos{\theta}$, where $\theta$ is the angle between the
line-of-sight and wave-number $\vec{k}$.  In deriving this expression
we have assumed $x_{\rm HI}\ll1$, as is observed following the
completion of reionization. The peculiar motion induced anisotropy is
dependent on cosmology through the evolution of growth factor $D$ via
$f=d\log{D}/d\log{(1+z)}$, and follows from application of redshift
space distortions (Kaiser 1987) to the 21cm PS. We assume $f=1$, which
is appropriate at high redshift. From equation~(\ref{DLAPSlin}) we
also find the spherically averaged PS
\begin{equation}
\label{DLAPSsph}
P_{21}(k) = 400\mbox{mK}\left(\frac{1+z}{7.5}\right)x_{\rm HI}^2\left[\langle b\rangle^2 + \frac{2}{3}\langle b\rangle + \frac{1}{5}\right] P(k).
\end{equation}
We note that in the case where $\langle b\rangle=1$ (which corresponds to a uniformly ionized IGM), we have 
\begin{equation}
P_{21}(k) = 400\mbox{mK}\left(\frac{1+z}{7.5}\right)\left[1.87\right] x_{\rm HI}^2P(k),
\end{equation}
which includes the enhancement of fluctuations by the factor of
$\langle(1+\mu^2)^2\rangle=1.87$ owing to peculiar velocities (Barkana
\& Loeb~2005). Importantly, the PS is sensitive only to the first
moment of $p(b)$. As a result, we do not need to calculate (or assume)
a functional form for this distribution in order to estimate the
constraints on $M_{\langle b\rangle}$.

Examples of spherically averaged power-spectra are plotted in
Figure~\ref{fig1} (thick grey lines) assuming values of $\langle
b\rangle$ corresponding to DLA halo masses of $M_{\langle
b\rangle}=10^{10}M_\odot$, $M_{\langle b\rangle}=10^{11}M_\odot$ and
$M_{\langle b\rangle}=10^{12}M_\odot$ (bottom to top), at each of the
redshifts $z=2.5$, 3.5, 4.5 and 5.5. We have assumed $x_{\rm HI}=0.02$
in these models corresponding to observations of HI density in DLAs at
high redshift (Prochaska et al.~2005). The wiggles at wave numbers
$k\sim0.1$Mpc$^{-1}$ are the baryonic oscillations that are familiar
to galaxy redshift surveys (Eisenstein et al.~2005). The models with
larger DLA host mass yield fluctuations with significantly more power,
owing to the larger galaxy bias. It is this variation in the power
that enables measurement of the DLA host mass.

\section{Sensitivity to fluctuations in 21cm emission from  DLAs}
\label{sensitivity}

As an example of the sensitivity of forthcoming low-frequency
telescopes to the fluctuations in 21cm emission after reionization, we
have estimated the signal-to-noise for the Murchison-Widefield Array
(MWA).  Calculations of the sensitivity to the 21cm PS for an
interferometer of this sort have been presented by a number of
authors. We follow the procedure outlined by McQuinn et al.~(2006),
drawing on results from Bowman, Morales \& Hewitt~(2006) for the
dependence of the array antenna density on radius $\rho(r)$.  The
uncertainty in a measurement of the PS per mode in the survey volume
has two separate components, due to the thermal noise of the
instrument ($\delta P_{\rm 21,N}$), and due to sample variance within
the finite volume of the survey ($\delta P_{\rm 21,SV}$). Since we are
dealing with discrete systems rather than a diffuse IGM, the latter
includes a Poisson component due to the finite sampling of each mode
and equals $\delta P_{21,SV}=P_{21}(k)[1+(\langle b \rangle ^2n_{\rm
DLA}P(k))^{-1}]$, in which we approximate the number density of
absorbing hosts by $n_{\rm DLA}\sim M\,dn_{\rm ST}/dM$ where $dn_{\rm
ST}/dM$ is the Sheth-Tormen~(2002) mass function of dark matter
halos. The noise due to sample variance therefore depends both on the
survey volume and on DLA host mass. We combine the above components to
yield the uncertainty on the estimate of the PS within a $k$-space
volume element $d^3k$
\begin{eqnarray}
\label{PSnoise}
\nonumber \Delta P_{21} &=& \left[\delta P_{\rm 21,SV} + \delta P_{\rm
21,N}\right]/\sqrt{N_{\rm c}}.
\end{eqnarray}
where the quantity $N_{\rm c} = 2\pi k^2\sin \theta
dkd\theta[\mathcal{V}/(2\pi)^3]$ denotes the number of modes observed
within $d^3k=2\pi
k^2\sin(\theta)dkd\theta$. In computing $N_{\rm c}$ we
assume symmetry about the polar angle and express the wave vector
$\vec{k}$ in components of its modulus $k$ and angle $\theta$ relative to
the line-of-sight. 

\begin{figure*}
\includegraphics[width=15cm]{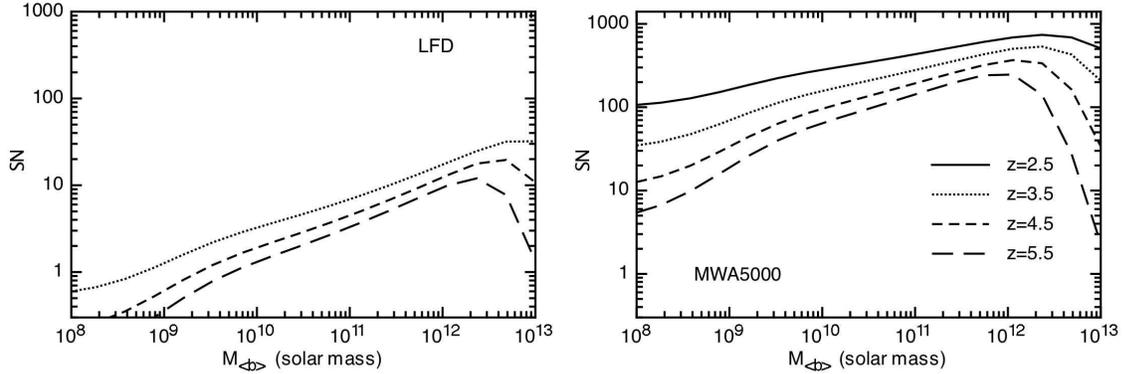} 
\caption{The integrated signal to noise for detection of the spherically averaged 21cm PS as a function of DLA mass. The cases of four different redshifts are shown for each of the MWA (left panel) and MWA5000 (right panel).}
\label{fig2}
\end{figure*} 

The contamination of foregrounds provides an additional source of
uncertainty in the estimate of the PS. McQuinn et al.~(2006) have
shown that it should be possible to remove the power due to
foregrounds to a level below the cosmological signal, provided that
the region of band-pass from which the PS is estimated is
substantially smaller than the total band-pass available. Following
the approximation suggested in McQuinn et al.~(2006), we set $N_{\rm
c}=0$ if $2\pi/k\, \cos(\theta)>\Delta D$, where $\Delta D$ is the
co-moving length corresponding to the line-of-sight distance over
which foregrounds can be removed.  The number of modes observed
depends on the volume of the survey, $\mathcal{V} = D^2\Delta
D(\lambda^2/A_{\rm tile})$, where $A_{\rm tile}$ is the total physical
surface area of an antenna, and $D$ is the co-moving distance to the
redshift of emission.

\subsection{Power-Spectrum Sensitivity of the MWA}

Estimates of the noise for detection of the spherically averaged PS
for the MWA are plotted in Figure~\ref{fig1} (thin solid lines). Note
that only redshifts $z\ga3.5$ are accessible with the MWA antennae
design. When complete, the MWA will comprise a phased array of 512
tiles (each tile will contain 16 cross-dipoles) distributed over an
area with diameter 1.5km.  To compute the noise on the PS we model the
antennae distribution as having $\rho(r)\propto r^{-2}$ with a maximum
radius of 750m and a finite density core of radius 18m. We assume a
1000hr integration on each of 3 fields, and a foreground removed
bandpass of $B=8$MHz within a total processed bandpass of 32MHz.

The sensitivity to the 21cm PS is dependent on both the sensitivity of
the telescope to a particular mode, and to the number of such modes in
the survey. The former is set by the effective collecting area
($A_{\rm e}$) of each antenna element (as well as the total number of
antennae), while the latter is sensitive to the total physical area
covered by each antenna (which we refer to as $A_{\rm tile}$). This
issue is discussed in more detail in Wyithe, Loeb \& Geil~(2008). In
computing the sensitivity we have assumed $A_{\rm e}\sim16
(\lambda^2/4)$m$^2$ and $A_{\rm tile}=16$m$^2$ (corresponding to the
design of the MWA). The combined uncertainties include the minimum $k$
cutoff due to foreground subtraction. Comparison of the noise estimate
with the expected 21cm signal shows that the MWA could detect the PS
at $3.5\la z\la5.5$ with high significance provided that the DLA hosts
are sufficiently massive (and therefore biased).

\subsection{Power-Spectrum Sensitivity of the MWA5000}

At values of $k\sim$ a few $\times10^{-1}$ Mpc$^{-1}$, the measurement
of the PS using the MWA will be limited by the thermal sensitivity of
the array, and so the signal-to-noise achievable in this regime will
be greatly enhanced by a subsequent generation of telescopes with
larger collecting area. As an example we consider a hypothetical
followup telescope to the MWA which would comprise 10 times the total
collecting area.  We refer to this followup telescope as the
MWA5000. The design philosophy for the MWA5000 would be similar to the
MWA, and we therefore assume antennae distributed as $\rho(r)\propto
r^{-2}$ with a diameter of 2km and a flat density core of radius 80m
(see McQuinn et al.~2006).  For MWA5000 we assume the antennae design
to be optimised at the redshift of observation (in which case we
assume $A_{\rm e}=A_{\rm tile}$). In Figure~\ref{fig1} we present
estimates for measurement uncertainty of the 21cm PS using MWA5000
(thin dashed lines). An MWA5000 would achieve a high signal-to-noise
detection of the PS, even in cases where the DLA hosts have a small
value of galaxy bias.

\begin{figure*}
\includegraphics[width=15cm]{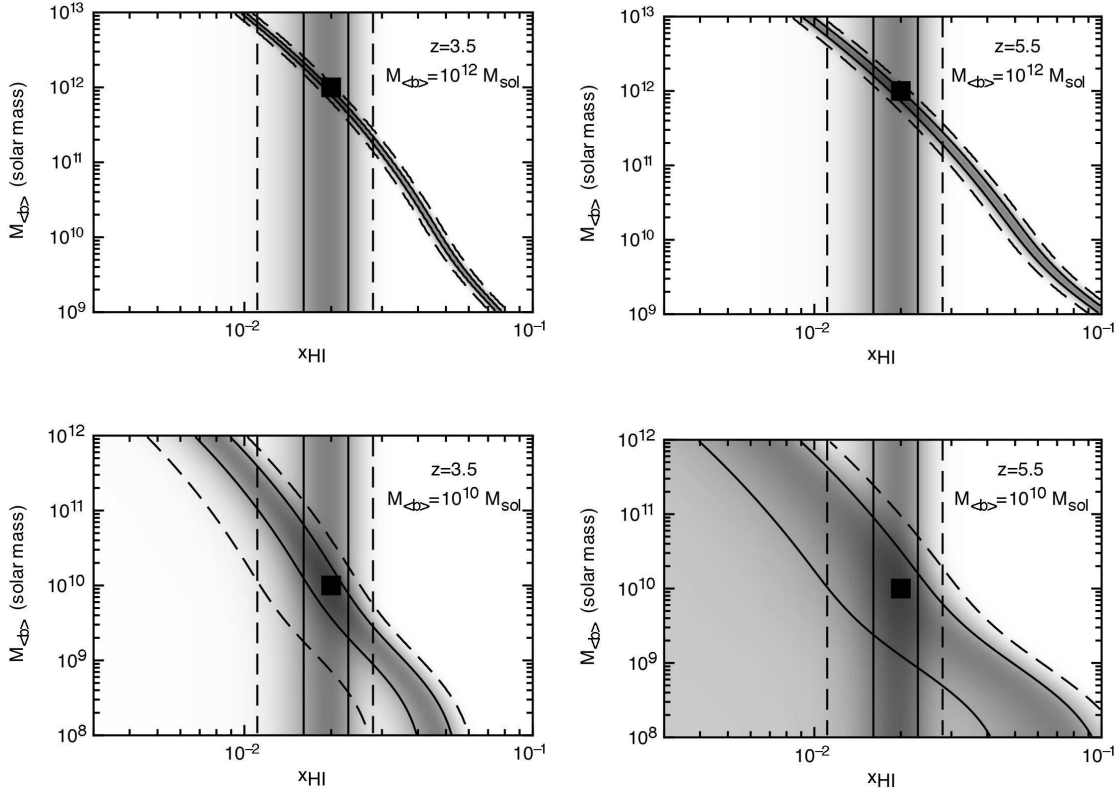} 
\caption{Constraints on DLA host mass and neutral fraction from a spherically averaged 21cm PS. The diagonal contours show the loci of points with likelihoods corresponding to 64\% (solid lines) and 10\% (dashed lines) of  the maximum, given a true model with DLA masses of $10^{12}M_\odot$ (upper panels) and $10^{10}M_\odot$ (lower panels) combined with a neutral fraction $x_{\rm HI}\sim0.02$ and observation with the MWA. The vertical lines show the 64\% and 10\% likelihood contours for neutral fraction based on existing observations of DLAs within a redshift bin of width $\delta z=0.5$.}
\label{fig3}
\end{figure*} 

\subsection{Dependence of $SN$ on DLA Host Mass}

In Figure~\ref{fig2} we show the integrated signal-to-noise ratio
($SN$) for detection of the spherically averaged PS as a function of
DLA host mass. The $SN$ improves with increasing host mass owing to
the increase in bias. Four different redshift cases are shown for each
of the MWA (left panel) and MWA5000 (right panel). Figure~\ref{fig2}
quantifies the results described in Figure~\ref{fig1}. The MWA can
detect the spherically averaged PS at a signal-to-noise of $\geq3$ for
DLA masses $M_{\langle b\rangle}\ga10^{11}M_\odot$, while the MWA5000
could detect the PS at a signal-to-noise of $\geq3$ for DLA masses as
small as $M_{\langle b\rangle}\sim10^{8}M_\odot$. In each case $SN$ is
limited at the highest masses by the Poisson noise introduced into the
PS by the discreteness of the DLA emission.

\section{Measurement of DLA Host Mass using Spherically Averaged 21cm Power Spectra}
\label{DLA}

In this section we consider the spherically averaged PS, which would
be the observable of choice for 21cm PS observations that achieve only
modest $SN$.  To estimate the potential for constraints on the DLA
host mass properties from the spherically averaged PS, we first
calculate the regions of parameter space $\vec{p}=(x_{\rm
HI},M_{\langle b\rangle})$ that are allowed around true models with
$\vec{p}_o$. We assume there is no uncertainty in the PS of matter
fluctuations, which would increase the uncertainty somewhat by
introducing distortions onto the PS via assumption of an incorrect
cosmology (Wyithe~2008). As part of this procedure, we construct
likelihoods
\begin{equation}
\ln{\mathcal{L}(\vec{p})} = -\frac{1}{2}\sum_{k}\left(\frac{P_{\Delta T}(k,\vec{p})-P_{\Delta T}^{\rm t}(k,\vec{p}_o)}{\Delta P^{\rm sph}_{\Delta T}(k)}\right)^2,
\end{equation}
where the sum is over bins of $k$, and $\Delta P^{\rm sph}_{\Delta T}$
is the spherically averaged uncertainty on measurement of the 21cm PS.
In Figure~\ref{fig3} we show the likelihood contours for combinations
of $x_{\rm HI}$ and $M_{\langle b\rangle} = M(\langle b\rangle)$ given
input models with $M_{\langle b\rangle}=10^{12}M_\odot$ (upper panels)
and $M_{\langle b\rangle}=10^{10}M_\odot$ (lower panels), combined
with $x_{\rm HI}=0.02$ and observation with the MWA (diagonal sets of
contours). The contours illustrate the degeneracy between $x_{\rm HI}$
and $M_{\langle b\rangle}$ in the resulting PS. Also shown is the
existing constraint on the neutral fraction from quasar absorption
line studies [vertical sets of contours showing the current
uncertainty within $\Delta z=0.5$ bins (Prochaska et al.~2005)]. The
figure illustrates how the neutral fraction constraint can be used to
break the degeneracy. It is clear from Figure~\ref{fig3} that the
constraint on $x_{\rm HI}$ will be the limiting factor for measurement
of the DLA host mass via the spherically averaged PS (in cases where
the host bias is large).

\begin{figure*}
\includegraphics[width=15cm]{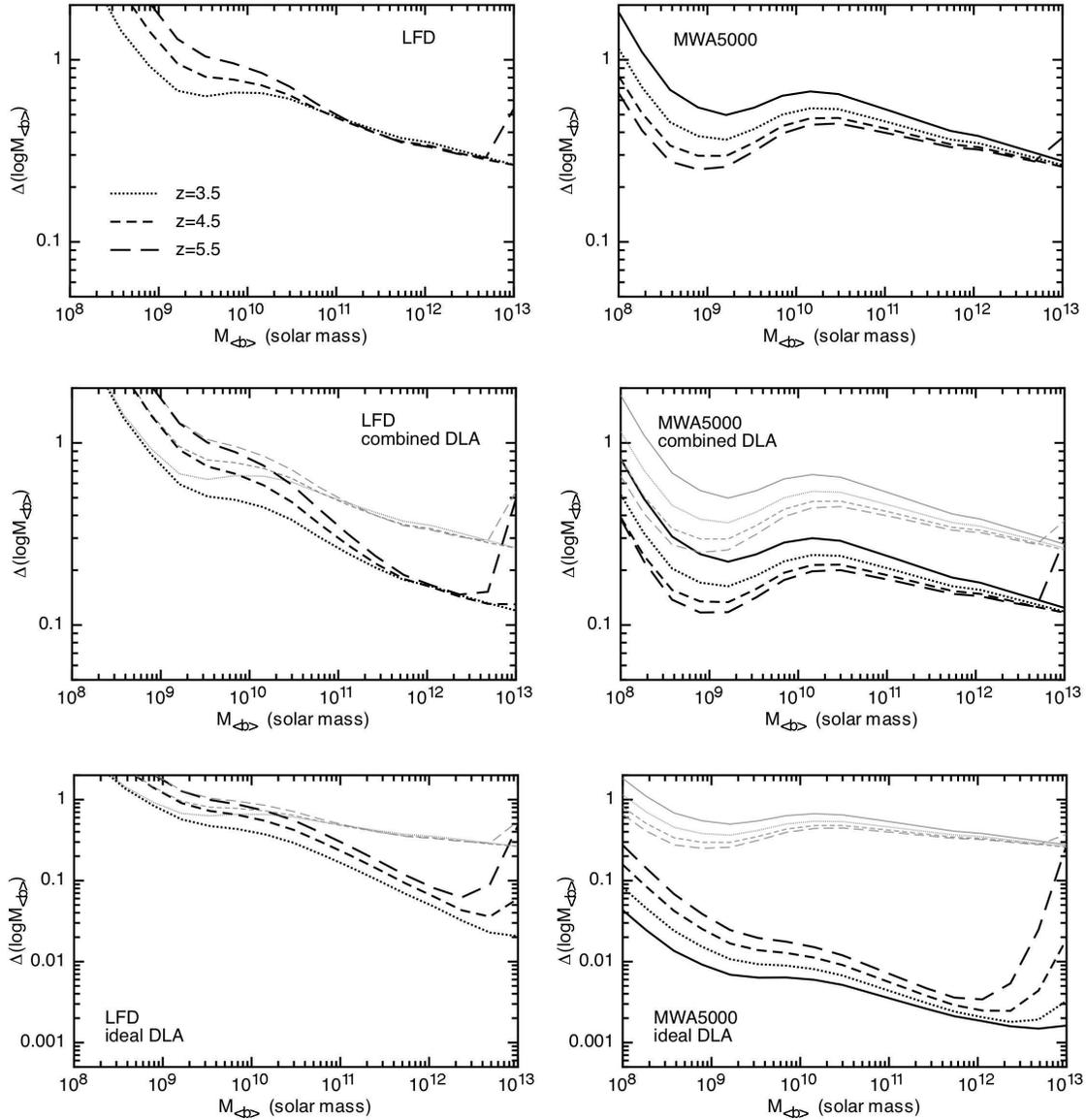} 
\caption{The uncertainty (in dex) on the mass of the DLA host achievable using a spherically averaged 21cm PS. The cases of different redshifts are shown for each of the MWA (left panels) and MWA5000 (right panels). {\em Upper panels:} The fractional uncertainty in $x_{\rm HI}$ was assumed to be $\Delta x_{\rm HI}/x_{\rm HI}=0.2$ (corresponding to one $\delta z=0.5$ bin in the SDSS study of Prochaska et al.~2005). {\em Central panels:}  The fractional uncertainty in $x_{\rm HI}$ was assumed to be $\Delta x_{\rm HI}/x_{\rm HI}=0.09$ (corresponding to a combination of all five $\delta z=0.5$ bins in the SDSS study, or a future survey with smaller error-bars).  {\em Lower panels:}  The fractional uncertainty in $x_{\rm HI}$ was assumed to be negligible. In the central and lower panels, the curves for $\Delta x_{\rm HI}/x_{\rm HI}=0.2$ are reproduced for comparison (thin grey curves). }
\label{fig4}
\end{figure*}

From equation~(\ref{DLAPSsph}) we see that measurement of the
spherically averaged PS with a signal to noise $SN$ yields a
fractional error on the combination $b_{\rm DLA}x_{\rm HI}$ with value
\begin{equation}
\label{eqnbx} \frac{\Delta (b_{\rm DLA}x_{\rm HI})}{b_{\rm DLA}x_{\rm
HI}} = \frac{1}{2 SN},  
\end{equation} 
where we have defined the quantity $b_{\rm DLA}=\sqrt{\langle
b\rangle^2+2/3\,\langle b\rangle+1/5}$. In deriving
equation~(\ref{eqnbx}) we have assumed the mass PS to have
negligible uncertainty.  The measurement of neutral fraction with
uncertainty $\Delta x_{\rm HI}$ is independent of the PS
(having been obtained via the column density in DLAs along the
line-of-sight towards high redshift quasars). The uncertainty in the
characteristic bias ($\Delta b_{\rm DLA}$) is therefore 
\begin{equation}
\label{uncertainty} \Delta b_{\rm DLA} = b_{\rm DLA}
\sqrt{\left(\frac{\Delta (b_{\rm DLA}x_{\rm HI})}{b_{\rm DLA}x_{\rm
HI}}\right)^2 + \left(\frac{\Delta x_{\rm HI}}{x_{\rm HI}}\right)^2}.
\end{equation} 
The corresponding uncertainty in $\langle b\rangle$ is
\begin{equation}
\Delta\langle b\rangle = \frac{d\langle b\rangle}{db_{\rm DLA}}\Delta b_{\rm DLA} = \frac{b_{\rm DLA}}{\langle b\rangle + 1/3}\Delta b_{\rm DLA},
\end{equation}
yielding the uncertainty in host mass (in dex)
\begin{equation} 
\label{mass_constraint} \Delta(\log{M_{\langle b\rangle}}) = \frac{d\log
M_{\langle b\rangle}}{d\langle b\rangle} \Delta \langle b\rangle, 
\end{equation} where
$d(\log{M_{\langle b\rangle}})/d\langle b\rangle$ is determined via equation~(\ref{bias}).

\begin{figure*}
\includegraphics[width=17cm]{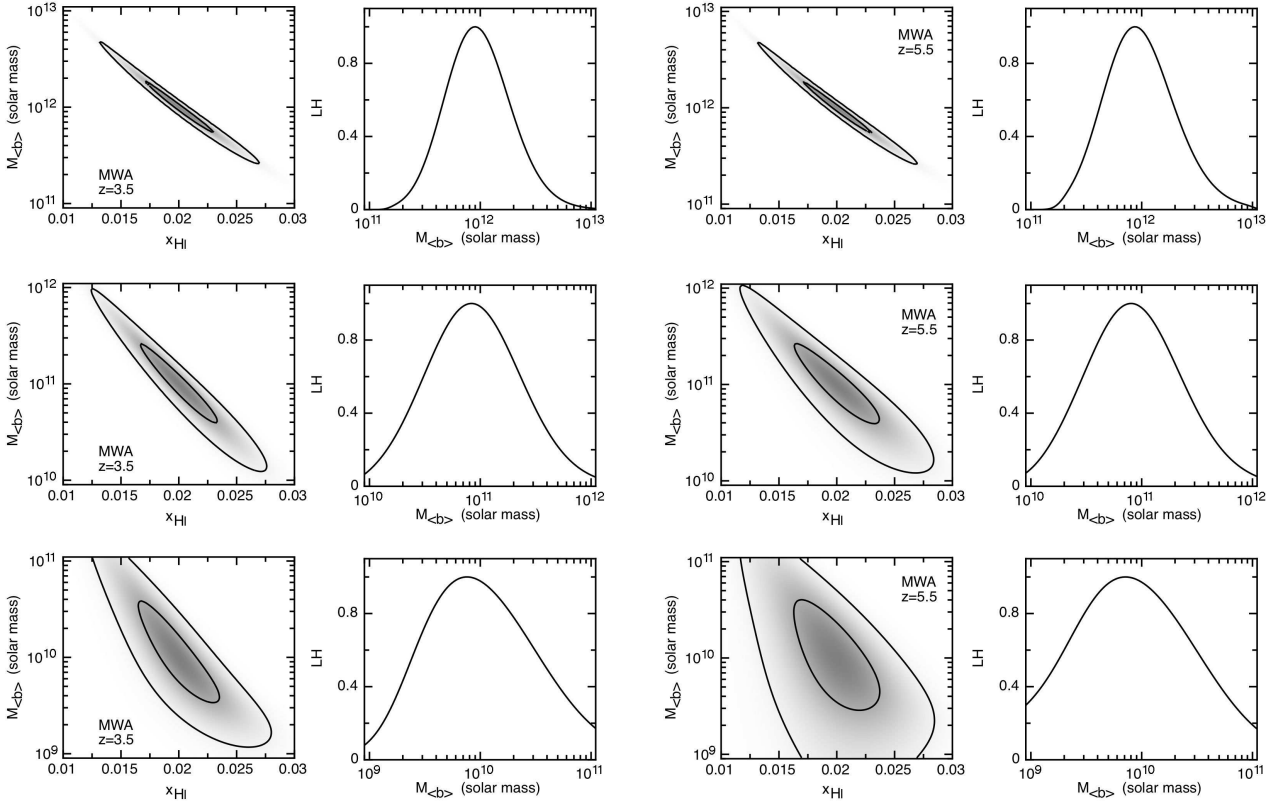} 
\caption{Constraints on the DLA host masses derived from measurement of the anisotropy of the 21cm PS. We assume observations using the MWA and 3 fields integrated for 1000 hours each. Results are shown at 2 redshifts, and for 3 values of the true host mass ($M_{\langle b\rangle}=10^{10}M_\odot$, $10^{11}M_\odot$ and $10^{12}M_\odot$; from bottom to top). We assume a true value of $x_{\rm HI}=0.02$. In each case likelihood contours are shown at 10\% and 64\% 
of the maximum for the parameter set $(x_{\rm HI},M_{\langle b\rangle})$. Also shown are the corresponding likelihoods for $M_{\langle b\rangle}$. We have  assumed a flat prior probability for $M_{\langle b\rangle}$, and a Gaussian prior probability for the neutral fraction with a variance of $\Delta x_{\rm HI}=0.2 x_{\rm HI}$. }
\label{fig5}
\end{figure*}

The values of $\Delta(\log{M_{\langle b\rangle}})$ that are obtained
via equation~(\ref{mass_constraint}) represent the dispersion of a
likelihood function $\mathcal{L}(\log{M_{\rm obs}}|\log{M_{\langle
b\rangle}})$ for observation of $\log{M_{\rm obs}}$ given a true value
$\log{M_{\langle b\rangle}}$. Following the observation of a 21cm DLA
PS the likelihood function could then be used to obtain an
a-posteriori measurement of the DLA host mass. Thus, in the instance
of a flat prior probability distribution for the logarithm of host
mass, the dispersion approximates the error (in dex) achievable for
the mass of the DLA host given the true mass $M_{\langle b\rangle}$.
The dispersion from equation~(\ref{mass_constraint}) is plotted in
Figure~\ref{fig4} as a function of host mass, at a range of redshifts,
and for different assumptions regarding the uncertainty in neutral
fraction and properties of the 21cm telescope used [the MWA (left
panels) and MWA5000 (right panels)]. The fractional uncertainty in
$x_{\rm HI}$ was assumed to be $\Delta x_{\rm HI}/x_{\rm HI}=0.2$
[upper panels; corresponding to one $\delta z=0.5$ bin in the SDSS DR3
study of Prochaska et al.~(2005)] and $\Delta x_{\rm HI}/x_{\rm
HI}=0.09$ [central panels; corresponding to a combination of all five
$\delta z=0.5$ bins in the Prochaska et al.~(2005) study]. With
current observations of DLA absorption systems, the host mass could be
determined to within a factor as small as 2-3, provided that the hosts
are massive.

\begin{figure*}
\includegraphics[width=17cm]{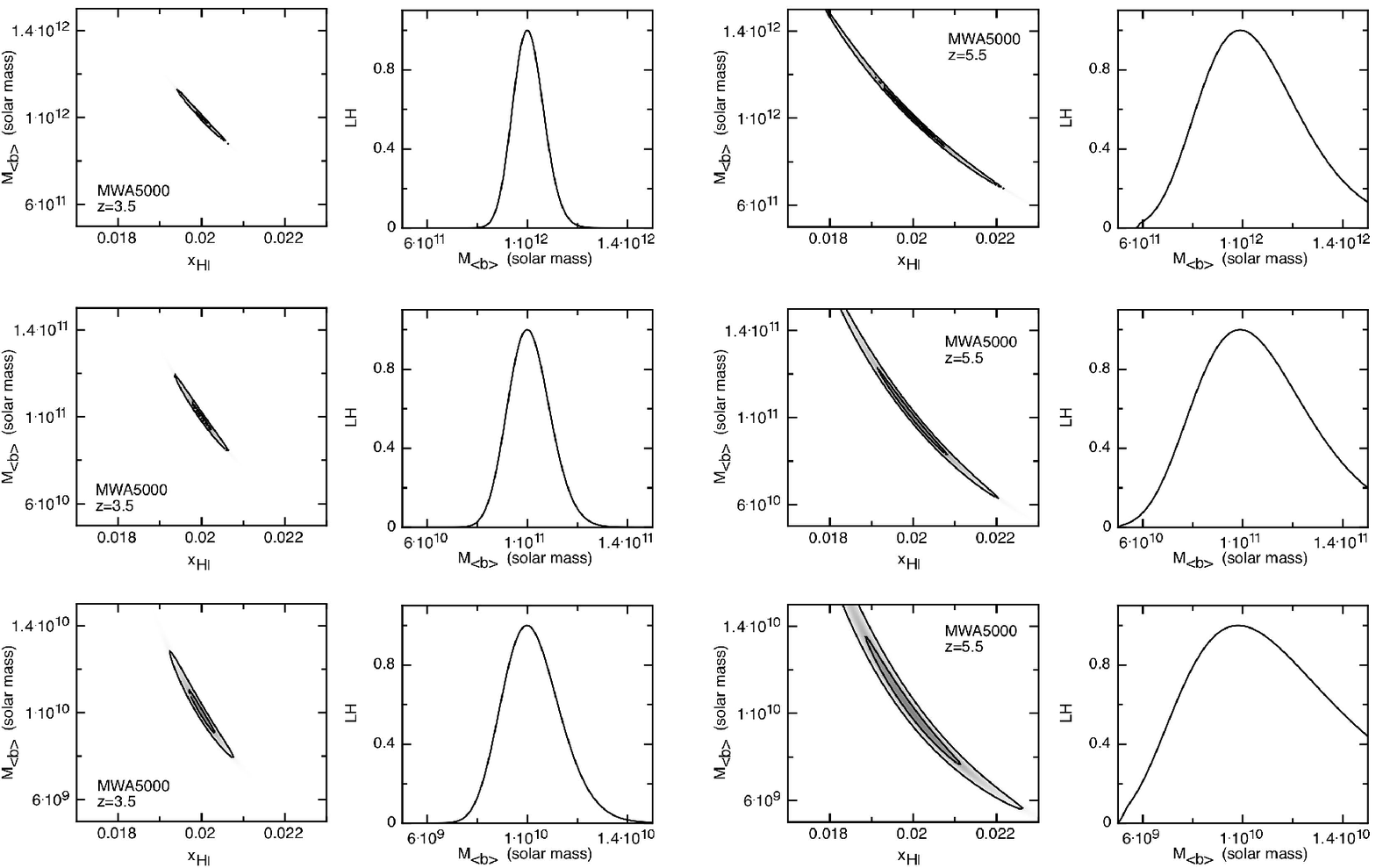} 
\caption{As per Figure~\ref{fig5}, but assuming observation using the MWA5000, and no prior constraint on $x_{\rm HI}$. }
\label{fig6}
\end{figure*}

It is apparent from equation~(\ref{uncertainty}) that the measurement
precision of the DLA clustering bias (and therefore the DLA host mass)
is limited by the uncertainty in neutral fraction once $SN\gg(x_{\rm
HI}/\Delta x_{\rm HI})/2\ga3-5$. Thus, with current observations of
quasar absorption systems, the DLA host mass estimate using a
spherically averaged PS would not be limited by the PS measurement
over a range of masses for the MWA. Moreover, the mass estimate would
not be limited by the PS measurement at any host mass using the
MWA5000. We therefore show additional results for the case where the
error in measurement of $\langle b\rangle$ is dominated by the
uncertainty in determination of the 21cm PS rather than the neutral
fraction. The resulting values of $\Delta(\log{M_{\langle b\rangle}})$
are shown in the lower panels of Figure~\ref{fig4}. The curves suggest
that the achievable accuracy would be limited by the determination of
$\Omega_{\rm HI}$ for host masses larger than $\sim10^{10}M_\odot$
using the MWA. In this most optimistic case the MWA5000 could measure
the characteristic DLA host mass to the level of a few percent over
the entire plausible mass range.

\section{Measurement of the DLA host mass using Anisotropy of the observed 21cm Power-Spectrum} 
\label{PSmu}

In the previous section we discussed the constraints that could be
placed on the DLA host mass via a spherically averaged 21cm PS. That
analysis broke the degeneracy between mass averaged neutral fraction
and the galaxy bias of DLA hosts using an independent measurement of
neutral fraction from the HI column density in quasar absorption
systems. In this section we use the peculiar velocity induced
anisotropy of the observed PS to break the degeneracy between bias and
neutral fraction using the fact that these quantities contribute
differently to the PS viewed at different angles relative to the
line-of-sight (equation~\ref{DLAPSlin}).

To estimate the potential for constraints on the DLA host mass, we
again calculate the region of parameter space $\vec{p}=(x_{\rm
HI},M_{\langle b\rangle})$ that is allowed around a true solution with
$\vec{p}_o$. We assume there is no uncertainty in the PS of matter
fluctuations, which would increase the uncertainty somewhat by
introducing distortions onto the PS via assumption of an incorrect
cosmology (Wyithe~2008). As part of this procedure, we construct
likelihoods
\begin{equation}
\ln{\mathcal{L}(\vec{p})} = -\frac{1}{2}\sum_{k,\mu}\left(\frac{P_{\Delta T}(k,\mu,\vec{p})-P_{\Delta T}^{\rm t}(k,\mu,\vec{p}_o)}{\Delta P_{\Delta T}(k,\mu)}\right)^2,
\end{equation}
where the sum is over bins of $k$ and $\mu$.

The results are shown in Figure~\ref{fig5}-\ref{fig6} for the MWA and
MWA5000 respectively using the observing strategy outlined in
\S~\ref{DLA}.  In each case likelihood contours are shown for the
parameter set $(x_{\rm HI},M_{\langle b\rangle})$ at two redshifts,
$z=2.5$ and 3.5, assuming a true value of $x_{\rm HI}=0.02$ and three
true values of host mass $M_{\langle b\rangle}=10^{10}M_\odot$,
$10^{11}M_\odot$ and $10^{12}M_\odot$. Also shown are the
corresponding likelihoods for $M_{\langle b\rangle}$.  We have assumed
a flat prior probability for $M_{\langle b\rangle}$. The $SN$ of the
MWA is not sufficiently large to break the degeneracy between bias and
neutral fraction. Therefore, as in the previous section we impose a
Gaussian prior probability on $x_{\rm HI}$, with a variance of $\Delta
x_{\rm HI}=0.2x_{\rm HI}$, corresponding to current uncertainty from
quasar absorption line studies (Prochaska et al.~2005).
Figure~\ref{fig5} shows that at the sensitivity of the MWA, the
constraints on the DLA host mass are very similar to those obtained
from the spherically averaged PS. As a result nothing is gained by
considering the angular dependence. On the other hand, the degeneracy
is broken at the $SN$ obtained by the MWA5000. As a result, we have
not imposed a prior probability on the neutral fraction from quasar
absorption studies in this case, and the allowed values of $x_{\rm
HI}$ in Figure~\ref{fig6} are derived directly from the 21cm PS, in
addition to $M_{\langle b\rangle}$. Figure~\ref{fig6} shows that the
MWA5000 (Figure~\ref{fig6}) would obtain accuracies of a few percent
on the neutral fraction, and a few 10s of percent in the DLA host
mass.

\section{Conclusions}
\label{conclusion}

Observations of DLA systems out to redshift $z\sim5$ show the density
parameter of HI to be $\Omega_{\rm HI}\sim10^{-3}$, indicating that
the mass averaged neutral hydrogen fraction remains at the level of a
few percent through most of cosmic history. These DLA systems will
produce 21cm intensity fluctuations, whose power-spectrum has an
amplitude that depends both on the total mass of neutral hydrogen
within the DLA systems ($\Omega_{\rm HI}$), and the masses of the DLA
host halos (through galaxy bias).  Since $\Omega_{\rm HI}$ is measured
via quasar absorption line studies, we show that measurement of the
spherically averaged 21cm power-spectrum amplitude at $2\la z\la5$
could be used to determine the DLA host mass. Using a telescope of
collecting area equal to the MWA, the DLA host mass could be
determined to within a factor as small as $\sim2$ (provided the hosts
were massive galaxies), with the accuracy limited by the determination
of $\Omega_{\rm HI}$ in this case.

In observations of the 21cm power-spectrum with high signal-to-noise,
the limitation of the quasar absorption line derived neutral fraction
could be removed by including the observed angular dependence due to
peculiar motions in the analysis. In this case the measurement of DLA
host mass would be limited by the sensitivity to the 21cm
power-spectrum, and future telescopes with larger collecting area
would significantly increase the precision (to within a few 10s of
percent) with which the DLA host mass could be determined.

The DLA host mass is currently estimated via clustering analyses of
absorption systems which are very rare, only being found in one of
every few quasar lines of sight. Thus measurement of the 21cm
power-spectrum, which measures the clustering of all DLAs in a large
volume, has the potential to greatly increase the precision with which
the DLA host mass is known. Since DLAs are thought to host the
majority of gas available for star formation, measurement of the DLA
host mass will be a valuable contribution to our understanding of
galaxy formation and the star formation history.

\bigskip

{\bf Acknowledgments} The author would like to thank Michael Murphy for helpful comments on the draft manuscript. This research was supported by the Australian Research
Council.

\newcommand{\noopsort}[1]{}

\label{lastpage}
\end{document}